\documentclass[times,5p,twocolumn]{elsarticle}

\usepackage{amsmath,amssymb,amsfonts}
\usepackage{algpseudocode}
\usepackage{algorithm}
\usepackage{bm}
\usepackage{booktabs}
\usepackage{cite}
\usepackage{float}
\usepackage{multirow}
\usepackage{layouts}
\usepackage{pifont}
\usepackage{subfig}
\usepackage{url}
\usepackage[usenames,dvipsnames]{xcolor}

\usepackage{wrapfig}
\usepackage{color}
\usepackage{capt-of}
\usepackage{tikz}
\usetikzlibrary{shapes,arrows,decorations.pathmorphing,backgrounds,fit,positioning,shapes.symbols,chains}

\def\subfigure{\subfloat}

\biboptions{numbers,sort,square}

\usepackage[figuresright]{rotating}

\begin{document}
\begin{frontmatter}





\title{TrendLearner: Early Prediction of Popularity Trends of User Generated Content} 

\author[fvdf]{Flavio Figueiredo\corref{cor}}
\ead{flaviov@dcc.ufmg.br}

\author[ufmg]{Jussara M. Almeida}
\ead{jussara@dcc.ufmg.br}

\author[ufmg]{Marcos A. Gon\c{c}alves}
\ead{mgoncalv@dcc.ufmg.br}

\author[ufmg]{Fabricio Benevenuto}
\ead{fabricio@dcc.ufmg.br}

\cortext[cor]{Corresponding author}
\address[ufmg]{Department of Computer Science, Universidade Federal de Minas Gerais\\
Av. Ant\^onio Carlos 6627, CEP 31270-010, Belo Horizonte - MG, Brazil. Phone: +55 (31) 3409-7541, Fax: +55 (31) 3409-5858}


\begin{abstract}
Predicting the popularity of user generated content  (UGC) is a valuable task to content providers, advertisers, as well as social media researchers. However, it is also a challenging task due to the plethora of factors that affect content popularity in social systems. Here, we focus on the problem of predicting the popularity {\it trend}  of a piece of  UGC (object)  {\it as early as possible}. Unlike previous work, we  explicitly address the inherent tradeoff between prediction accuracy and remaining interest in the object after prediction, since, to be useful, accurate predictions should be made {\it before}  interest has exhausted. Given the heterogeneity in popularity dynamics across objects, this tradeoff has to be solved on a per-object basis, making the prediction task harder. We tackle this problem with a novel two-step learning approach in which we: (1)  extract popularity trends from previously uploaded objects, and then (2) predict trends for newly uploaded content. Our results for YouTube datasets show that our classification effectiveness, captured by F1 scores, is 38\% better than the baseline approaches. Moreover, we achieve these results with up to 68\% of the views still remaining for 50\% or 21\% of the videos, depending on the dataset.

\end{abstract}

\end{frontmatter}

\section{Introduction}
\label{sec:intro}

The success of Internet applications based on user generated content (UGC)\footnote{YouTube, Flickr, Twitter, and so forth}
has motivated questions such as: How does content popularity  evolve over time?
What is the potential popularity  a piece of content will
achieve after a given time period? How can we predict popularity
evolution of a particular piece of UGC?
For example, from a system perspective, accurate popularity predictions can be exploited to build more cost-effective content organization and delivery platforms (e.g., caching systems, CDNs). They can also drive the design of better analytic tools, a major segment nowadays~\citep{Leskovec2011,Zeng2010}, while online advertisers may benefit from them to  more effectively place contextual advertisements.
From a social perspective, understanding issues related to popularity prediction can be used to  better understand the human dynamics of consumption. Moreover, being able to predict popularity on an automated way is crucial for marketing campaigns (e.g. created by activists or politicians),  which  increasingly often use the Web to  influence public opinion.

\begin{figure*}[t]
   \centering
    \includegraphics[scale=1.7]{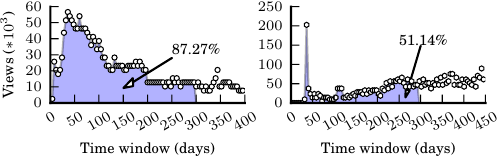}
    \caption{Popularity Evolution of Two YouTube Videos.}
    \label{fig:ex}
\end{figure*}

{\bf Challenges: } However, predicting the popularity of a piece of content, here referred to as an {\it object},  in a social system is a  very challenging task. This is mostly due to the various phenomena affecting the popularity prediction of social media -- which were observed on the datasets we use (as well as others)~\citep{Figueiredo2011,Matsubara2012,Yu2015} -- as well as the diminishing interesting in objects over time, which implies that popularity predictions must be timely to capture user interest and be useful in real work settings. Both challenges can be summarized as follows:
\begin{enumerate}
    \item Due to the easiness with which UGC can be created, many factors can affect an object's popularity. Such factors include, for instance, the  object's content, the social context in which it is inserted (e.g., social neighborhood or influence zone of the object's creator), the mechanisms used to access the content (e.g., searching, recommendation,  top-lists), or even an external factor, such as a hyperlink to the content in a popular blog or website. These factors can cause spikes in the surge of interest in objects, as well as information propagation cascades which affect the popularity trends of objects.
    \item To be useful in a real scenario, a popularity prediction approach must identify popularity trends {\it before the user interest in the object has severely diminished.} To illustrate this point, Figure \ref{fig:ex} shows the popularity evolution of two YouTube videos: the video on the left receives more than $80\%$ (shaded region) of all views received during its  lifespan in the first 300 days since upload, whereas the other video  receives only about half of its total views in the same time frame. If we were to monitor each video for 300 days, most potential views of  the first video would be lost. In other words, not all objects require the same monitoring period, as assumed by   previous work, to produce accurate predictions: for some objects, the prediction can be made earlier. Thus, the tradeoff should be solved on a  {\it per-object} basis, which implies that determining the duration of the monitoring period that leads to a good solution of the tradeoff for {\it each object} is part of the problem.
\end{enumerate}

These challenges set UGC objects apart from more traditional web content. For instance, news media~\citep{Castillo2014} tends to have clear definitions of monitoring periods, say predicting the popularity of news after one day using information from the first hour after upload. This is mostly due to the timely nature of the content, which is reflected in the popularity trends usually followed by news media~\citep{Figueiredo2011} -- interest is usually concentrated in a peak window (e.g., day) and dies out rather quickly.
Thus, mindful of the challenges above, we here tackle the problem of UGC popularity {\it trend} prediction. That is, we focus on the (hard) task of predicting popularity {\it trends}. Trend prediction can help determining, for example, if  an object will follow a viral pattern (e.g.,  Internet memes) or will continue to gain attention over time (e.g.,  music videos for popular artists). Moreover, we shall also show that, by knowing popularity trends {\it beforehand}, we can improve the accuracy of  models for predicting popularity measures (e.g., hits). Thus, by focusing on predicting trends, we fill a gap  in current research since no previous efforts has effectively predicted the popularity {\it trend} of UGC taking into account  challenges (1) and (2).

We should stress that one key aspect distinguishes our work from previous efforts to predict popularity~\citep{Castillo2014,Lerman2010,Yin2012,Szabo2010,Pinto2013,Ahmed2013} --
 we explicitly address the inherent tradeoff between prediction accuracy and how early the prediction is made,  assessed in terms of the remaining interest in the content after prediction. All previous popularity prediction efforts considered fixed monitoring periods for all objects, which is given as input. We refer to this problem as {\it early prediction}\footnote{We also point out that an earlier, much simpler, variant of our approach, which did not focus on early predictions, was first place on two out of three prediction tasks of the 2014 ECML/PKDD Predictive Analytics Challenge for News Content~\citep{Figueiredo2014}\footnote{http://sites.google.com/site/predictivechallenge2014/},  reflecting the quality/effectiveness of our proposal.}.

 In terms of applications, knowing that an object will be popular early on can help advertisers to plan out specific revenue models~\citep{Gill2013}. Such knowledge can also help out on geographic content sharding~\citep{Duong2013} for better content delivery. On the other hand, being aware that an object will not be popular at all, as early as possible, allow  low access content to be tiered down to lower latency servers/geographic regions, whereas advertisers can use this knowledge to avoid bidding for ads in such content (since they will not generate revenue). Another example are search engines rankings based on predictions~\citep{Radinsky2012}. Knowing that a content is becoming popular can help out in generating better rankings to user queries. However, if we have evidence (based on the trend and remaining interest) that such content is losing popularity (e.g., timely content that users can lose interest over time), such contents may be of less interest to the user. Finally, early prediction is of utmost importance to content producers --  knowing whether a piece of content will be follow a certain trend can help in their promotion strategies and in the creation of new content.


{\bf TrendLearner:}
We tackle this problem  with a novel two-step combined learning approach. First, we identified popularity trends, expressed by popularity timeseries, from previously uploaded objects. Then, we combine novel time series classification algorithms with object features for predicting the trends of new objects. This approach is motivated by the intuition that it might be easier to identify the popularity trend of an object if one has a set of possible trends as basis for comparison. More important, we propose a new trend classification approach, namely TrendLearner, that tackles the aforementioned  tradeoff between prediction accuracy and remaining interest after prediction on a per-object basis. The idea here is to monitor newly uploaded content on an online basis to determine, for each monitored object, the earliest point in time when prediction confidence is deemed to be good enough (defined by input parameters), producing, as output, the probabilities of each object belonging to each class (trend). Moreover, unlike previous work, TrendLearner also combines the results from this classifier (i.e., the probabilities) with a set of object related features \citep{Figueiredo2011}, such as category and incoming links, building an ensemble learner.

To evaluate our method, we use, in addition to traditional classification metrics (e.g., Micro/Macro F1),  two newly defined metrics, specific for the problem: (1) remaining interest (RI), defined as the fraction of all views (up to a certain date) that remain after the prediction, and (2)  the correlation between the total views and the remaining interest. While the first metric measures the potential future viewership of the objects, the second one estimates whether there is any bias towards more/less popular objects.


In sum, our main contributions include a novel popularity trend classification method that considers multiple trends, called TrendLearner. The use of TrendLearner can improve the prediction of popularity metrics (e.g., number of views). Improvements over state-art-method are significant, being around 33\%, at least.


The rest of this article is organized as follows. Next section discusses related work. We state our target problem in Section~\ref{sec:meth}, and present our approach to solve it in Section \ref{sec:ourapproach}. We introduce the metrics and  datasets used to evaluate  our approach in Section~\ref{sec:methodology}. Our main experimental results are discussed in Section \ref{sec:res}. Section~\ref{sec:disc} offers conclusions and directions for future work.

\section{Related Work} \label{sec:related}

Popularity evolution of online content has been the target of several studies.  Several previous efforts aimed at developing models to predict the popularity of a piece of content at a given future date. In \citep{Lerman2010}, the authors developed stochastic user behavior models to predict the popularity of Digg's stories based on early user reactions to new content and aspects of the website design. Such models are very specific to Digg features,  and are not general enough for different kinds of UGC. Szabo and Huberman proposed a linear regression method  to predict the  popularity of YouTube and Digg content from early measures of user accesses~\citep{Szabo2010}.  This method has been recently extended and improved with the use of multiple features~\citep{Pinto2013}. Castillo {\it et. al.}~\citep{Castillo2014} used a similar approach as~\citep{Szabo2010} to predict the popularity of news content.

Out of these previous efforts, most authors focused on variations of Linear Regression based methods to predict UGC popularity~\citep{Lee2010,Pinto2013,Castillo2014}. In the context of search engines, Radinsky {\it et al.} proposed Holt-Winters linear models to predict future popularity, seasonality and the bursty behavior of queries~\citep{Radinsky2012} . The models capture the   behavior of a population of users searching on the Web for a specific query,  and are trained for each individual time series. We note that none of these prior efforts focused on the problem of {\it predicting popularity trends}. In particular, those focused on UGC popularity prediction assumed a fixed monitoring period for all objects, given as input, and did not explore the trade-off between prediction accuracy and remaining views after prediction.

Other methods exploit epidemic modeling of UGC popularity evolution. Focusing on content propagation within an OSN, Li {\it et al.} addressed video popularity prediction within  a single (external) OSN (e.g., Facebook)~\citep{Li2013}. Similarly, Matsubara {\it et. al.}~\citep{Matsubara2012} created a unifying epidemic model for the trends usually found in UGC. Such a model can be use for tail forecasting, that is, predictions after the peak time window. Again, none of these methods focus neither on trend predictions or on early predictions as we do. Also, tail-part forecasting is very limited when the popularity of an object may exhibit multiple peaks~\citep{Hu2014,Yu2015}. By focusing on a two step trend identification and prediction approach, combined with a non-parametric distance function, TrendLearn can overcome these challenges.



Chen {\it et al.}~\citep{Chen2013} propose to predict whether a tweet will become a trending topic by applying a binary  classification model (trending versus non-trending), learned from a set of objects from each class. We here propose a more general approach to detect {\it multiple} trends (classes), where trends are first {\it automatically}  learned from a training set. It is also important to note that our solution complements the one by Jiang {\it et al.}~\citep{Jiang2014}, which focused on predicting when a video will peak in popularity. Finally, our solution also exploits the concept of shapelets~\citep{Ye2011} to reduce the classification time complexity, as we show in Section \ref{sec:ourapproach}.

We also mention some other efforts to detect trending topics in various domains.  Vakali \textit{et al.} proposed a cloud-based framework for detecting trending topics on Twitter and blogging systems~\citep{Vakali2012}, focusing particularly on implementing the framework on the cloud, which is complementary to our  goal. Golbandi {\it et al.}~\citep{Golbandi2013}  tackled trend topic detection for search engines. Despite the similar  goal, their solution applies to a  very different domain, and thus focuses on different elements (query terms) and  uses different techniques (language models) for prediction. 


\begin{table*}[t]
\centering
\caption{Comparison of TrendLearner with other approaches}
\begin{tabular}{@{}lcccc@{}} 
\toprule
& Trend Identification & Trend Prediction & Views Prediction & Early Prediction  \\
\midrule
Trending Topics Prediction \\
~\citep{Chen2013,Vakali2012} & & \checkmark (Binary only) & & \\
\\
\midrule
Linear Regression~\citep{Pinto2013} \\
~\citep{Szabo2010}         & & & \checkmark   & \\
~\citep{Castillo2014} \\
\midrule
Holt-Winters \\
~\citep{Radinsky2012}     & & & \checkmark & \\
\\
\midrule
Epidemic Models \\
~\citep{Li2013,Matsubara2012} & & & \checkmark & \\
\\
\midrule
TrendLearner  &  \checkmark  &  \checkmark  &  \checkmark  &  \checkmark  \\
\bottomrule
\label{tab:salesmat}
\end{tabular}
\end{table*}

Table~\ref{tab:salesmat}  summarizes the key functionalities of the aforementioned approaches as well as of our new TrendLearner method.
In sum, to our knowledge, we are the first to tackle the inherent challenges of predicting UGC popularity (trends and metrics) as early and accurately as possible, on a per-object basis, recognizing that different objects may require different monitoring periods for accurate predictions. More important, the challenges we approach with TrendLearner (i.e. predicting trends also tackling the  tradeoff between prediction accuracy and remaining interest after prediction on a per-object basis) are key to leverage popularity prediction towards practical scenarios and deployment in real systems.

\section{Problem Statement} \label{sec:meth}

The early popularity trend prediction problem can be defined as follows. Given a training set of previously monitored user generated objects (e.g., YouTube videos or tweets), $D^{train}$, and a test set of newly uploaded objects $D^{test}$, do: (1) extract popularity trends from  $D^{train}$; and (2) predict a trend  for each object in $D^{test}$ as early and accurately as possible, particularly before user interest in such content has significantly decayed. User interest can be expressed as the  fraction of all  potential views a new content will receive until a given point in time (e.g., the day when the object was collected). Thus, by predicting as early as possible the popularity trend of an object,  we aim at maximizing the fraction of views that still remain to be received {\it after prediction}.  Determining the earliest point in time when prediction can be made with reasonable accuracy is an inherent challenge of the early popularity   prediction problem, given that it must be addressed on a per-object basis. That is, while later predictions can be more accurate, they would imply a reduction of the remaining interest in the content. 

In particular, we here treat the above problem as a trend-extraction one combined with as a multi-class classification task. The popularity trends automatically extracted from $D^{train}$ (step 1) represent the classes into which objects in $D^{test}$ should be grouped (step 2). Trend extraction is performed using a time series clustering algorithm~\citep{Yang2011}, whereas prediction is a classification task. For the sake of clarity, we shall make use of the term ``class'' to refer to both clusters and classes. 

\begin{table*}[t]
\centering
\caption{Notation. Vectors ($\mathbf{x}$) and matrices ($\mathbf{X}$), in bold, are differentiated by lower and upper cases. Streams ($\mathbf{\hat{x}}$) are differentiated by the hat accent ($\,\bm{\hat{}}\,$).  Sets ($D$)  and variables ($d$)  are shown  in regular upper and lower case letters,  respectively.} 
\begin{tabular}{lll}
\toprule
Symbol & Meaning & Example \\
\midrule
    $D$ & dataset of UGC content & YouTube videos \\
    $D^{train}$ & training set & - \\
    $D^{test}$ & testing set & - \\
    $d$ & a piece of content or object & video \\
    $D_i$ &  class/trend i & - \\
    $\mathbf{c}_{D_i}$ & centroid of class i & - \\
    $\mathbf{s}_d$ &  time series vector for object $d$ & $\mathbf{s}_d = <p_{d,1}, \cdots, p_{d,n}>$ \\
    $\mathbf{\hat{s}}_d$ & time series stream for object $d$ & $\mathbf{\hat{s}}_d = <p_{d,1}, , \cdots$ \\
    $p_{d,i}$ & popularity of  $d$ at i-th window & number of views \\
    $\mathbf{s}_d^{[i]}$ & index operator & $<7, 8, 9>^{[2]} = 8$ \\
    $\mathbf{s}_d^{[i:j]}$ & slicing operator & $<7, 8, 9>^{[2:3]} = <8, 9>$ \\
    $\mathbf{S}$ & matrix with set of time series& all time series \\
\bottomrule
\end{tabular}
\label{tab:notation}
\end{table*}

Table~\ref{tab:notation} summarizes the notation used throughout the paper. Each object  $d \in D^{train}$ is represented by 
 an $n$-dimensional time series vector  $ \mathbf{s}_d = <p_{d,1}, p_{d,2}, \cdots, p_{d,n}>$,  where $p_{d,i}$ is the popularity (i.e., number of views) acquired by    $d$  {\it during} the $i^{th}$ time window after its upload. Intuitively, the duration of a time window $w$ could be a few  hours, days, weeks, or even months. Thus, vector $\mathbf{s}_d$ represents a time series of the popularity of a piece of content measured at time intervals of duration $w$ (fixed for each vector). New objects in $D^{test}$ are represented by streams, $\mathbf{\hat{s}}_d$, of potentially infinite length ($\mathbf{\hat{s}}_d = <p_{d,1}, p_{d,2}, \cdots$). This captures the fact that our trend prediction/classification method is based on monitoring each test object on an online basis, determining when a prediction with acceptable confidence can be made (see Section \ref{subsec:TrendLearner}). Note that a vector can be seen as a contiguous subsequence of a stream. Note also that the complete dataset is referred to as $D = D^{train} \bigcup D^{test}$.  

 \section{Our Approach} \label{sec:ourapproach}

We here present our solution to the early popularity trend prediction problem. We  introduce our trend extraction approach  (Section \ref{subsec:trendextraction}), present  our novel trend classification method, TrendLearner (Section \ref{subsec:TrendLearner}), and  discuss practical issues related to the joint use of both techniques (Section \ref{sec:together}). 

\subsection{Trend Extraction} \label{subsec:trendextraction}

To extract temporal patterns of popularity evolution (or trends) from objects in $D^{train}$, we employ a time series clustering algorithm  called K-Spectral Clustering (KSC)~\citep{Yang2011}\footnote{We have implemented a parallel version of the KSC algorithm which is available at \url{http://github.com/flaviovdf/pyksc}. The repository also contains the TrendLearner code}, which groups time series  based on the {\it shape} of the curve. 
To group the time series, KSC defines the following distance metric to capture the similarity between two  time series $\mathbf{s}_d$ and  $\mathbf{s}_{d'}$ with scale and shifting invariants:

\begin{equation} \label{eq:dist}
dist(\mathbf{s}_d, \mathbf{s}_{d'}) = \displaystyle\min{\alpha, q} \quad \frac{||\mathbf{s}_d - \alpha\mathbf{s}_{d'(q)}||}{||\mathbf{s}_d||},
\end{equation}

\noindent where $\mathbf{s}_{d'(q)}$ is the operation of shifting the time series $\mathbf{s}_{d'}$ by $q$ units and $||\cdot||$ is the $l_2$ norm\footnote{The $l_2$ norm of a vector $\mathbf{x}$ is defined as $||\mathbf{x}|| = \sqrt{\sum_{i=1}^n  x_i^2}$.}.  For a fixed $q$, there exists an exact solution for $\alpha$ by computing the minimum of $dist$, which is: $\alpha = \frac{\mathbf{s}_{d}^{T}\mathbf{s}_{d'(q)}}{||\mathbf{s}_{d'}||}.$ In contrast, there is no simple way to compute shifting parameter $q$. Thus, in our implementation of KSC, whenever we measure the distance between two series, we search for the optimal value of  $q$ considering all integers in the range $(-n,n)$\footnote{Shifts are performed in a rolling manner, where elements at the end of the vector return to the beginning. This maintains the symmetric nature of $dist(\mathbf{s}_d, \mathbf{s}_{d'})$.}. 

Having defined a distance metric, KSC  is mostly a direct translation of the K-Means algorithm~\citep{Coates2012}. Given a number of trends $k$ to extract and the set of time series,  it works as:

\vspace{0.2cm}
\noindent 1. The time series are uniformly distributed to $k$ random classes;\vspace{0.2cm}


\noindent 2. Cluster centroids are computed based on its  members. In K-Means based algorithms, the goal is to find centroid $\mathbf{c}_{D_i}$ such that $\mathbf{c}_{D_i} = arg,min_{\bm{c}} \sum_{\mathbf{s}_d \in D_i} dist(\mathbf{s}_d, \bm{c})^2$. We refer the reader to the original KSC paper for more details on how to find $\mathbf{c}_{D_i}$~\citep{Yang2011};
\vspace{0.2cm}
\noindent 3. For each time series vector $\mathbf{s}_d$,  object $d$ is assigned to the nearest centroid based on  metric $dist$;\vspace{0.2cm}
\noindent 4. Return to step 2 until convergence, i.e., until all objects remain within the same class in step 3. \vspace{0.2cm}
Each centroid  defines the trend that objects in the class (mostly) follow. 

Before introducing our trend classification method, we make the following observation that is key to support the design of the proposed approach: each trend, as defined by a centroid, is conceptually equivalent to the notion of {\it time series shapelets} \citep{Ye2011}. A shapelet is informally defined as a time series subsequence that is in a sense maximally representative of a class.
 As argued in \citep{Ye2011}, the distance to the shapelet can be used to classify objects with more accuracy and much faster than state-of-the-art classifiers. Thus, by showing that a centroid is a shapelet, we choose to classify a new object  based only on the distances between the object's popularity time series up to a monitored time and each trend.

 This is one of the  points where our approach differs from the method proposed in \citep{Chen2013}, which uses the complete $D^{train}$ as reference series, classifying an object  based on the distances between its time series and {\it all} elements of each class.  Given $|D^{train}|$ objects in the training set and $k$ trends (with $k << |D^{train}|$), our approach is faster by a factor of $\frac{|D^{train}|}{k}$.

 {\bf Definition:} {\it For a given class $D_i$, a shapelet $\mathbf{c}_{D_i}$ is a time series subsequence such that: } (1) $dist(\mathbf{c}_{D_i}, \mathbf{s}_d) \leq \beta,\forall \mathbf{s}_d \in D_i$;  and (2) $dist(\mathbf{c}_{D_i}, \mathbf{s}_{d'}) > \beta, \forall \mathbf{s}_{d'} \notin D_i$, where $\beta$ is defined as an optimal distance for a given class. With this definition, a shapelet can be shown to maximize the information gain of a given class~\citep{Ye2011}, being thus the most representative time series of that class.

We argue that, by construction, a  centroid produced by KSC is a shapelet with $\beta$ being the  distance from the centroid to the time series within the class that is furthest away from its centroid. Otherwise, the time series that is furthest away would belong to a different class, which contradicts the KSC algorithm. This is an intuitive observation. Note that a centroid is a shapelet {\em only} when using  K-Means based   approaches, such as KSC,  to define class labels.  In the case of learning from already labeled data a shapelet finding algorithms~\citep{Ye2011} should be employed.
 
\subsection{Trend Prediction} \label{subsec:TrendLearner}

Let $D_i$ represent class  $i$, previously learned from $D^{train}$. 
Our task now is to create a classifier that  correctly determines the class of  a new object  as early as possible. We do so by monitoring the popularity acquired by each object $d$ ($d \in D^{test}$) since its upload on successive time windows. As soon as we can state that $d$ belongs to a  class  with  {\it acceptable confidence}, we stop monitoring it and report the prediction. The heart of this approach is in detecting {\it when} such statement can be made.

\subsubsection{Probability of an Object Belonging to a Class} \label{subsec:prob}

\begin{figure}[ht]
    \centering
    \includegraphics{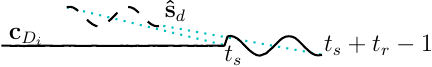}
    \caption{Example of alignment of time series  (dashed lines) for probability computation.}
    \label{fig:exprob}
\end{figure}

Given a monitoring period defined by $t_r$ time windows, our trend prediction is fundamentally based on the distances between the subsequence of the stream $\mathbf{\hat{s}}_{d}$ representing $d$'s popularity curve from its upload until $t_r$,  $\mathbf{\hat{s}}_{d}^{[1:t_r]}$, and the centroid of each class.  To respect shifting invariants, we consider all possible starting windows $t_s$ in each centroid time series when computing distances. That is, given a centroid $\mathbf{c}_{D_i}$, we consider all values from 1 to $ |\mathbf{c}_{D_i}| - t_r$, where $ |\mathbf{c}_{D_i}|$ is the number of time windows in $\mathbf{c}_{D_i}$.
Specifically,  the probability that  a new object $d$   belongs to  class $D_i$, given $D_i$'s centroid, the monitoring period $t_r$ and a starting window $t_s$, is:

\begin{eqnarray} \label{eq:prob}
    p(\mathbf{\hat{s}}_{d}\in D_i \mid \mathbf{c}_{D_i}; t_r, t_s) \propto 
    exp(-dist(\mathbf{\hat{s}}_{d}^{[1:t_r]}, \mathbf{c}_{D_i}^{[t_s: t_s + t_r - 1]}))
\end{eqnarray}

\noindent where $[x$:$y]$ ($x \leq y$) is a moving window slicing operator (see Table \ref{tab:notation}). As in \citep{Chen2013,Pinto2013,Coates2012}, we assume that probabilities are inversely proportional to the  exponential function of the distance between both  series, given by function $dist$ (Equation \ref{eq:dist}), normalizing them afterwards to fall in the 0 to 1 range (here omitted for simplicity). Figure~\ref{fig:exprob} shows an illustrative example of how both time series would be aligned for probability computation\footnote{In case   $|\mathbf{c}_{D_i}| < |\mathbf{\hat{s}}_{d}^{[1:t_r]}|$,   we try all possible alignments of $\mathbf{c}_{D_i}$ with $\mathbf{\hat{s}}_{d}^{[1:t_r]}$.}. That is, for time series of different lengths, we slice a consecutive range of the largest time series so that it has the size of the smallest one. Every slice possible is considered (starting from 1 to $|\mathbf{\hat{s}}_{d}^{[1:t_r]}|$) and we keep the slice with the smallest distance when computing probabilities.

With Equation \ref{eq:prob}, we could build a classifier that simply picks the class with highest probability. But  this would require  $t_s$ and $t_r$ to be fixed.  As shown in Figure~\ref{fig:ex},  different time series may need different monitoring periods (different values of $t_s$ and $t_r$), depending on the required confidence. 

Instead, our approach is to monitor an object for successive time windows (increasing $t_r$), computing the probability of it belonging to each class at the end of each  window. We stop when the class with maximum probability exceeds a class-specific threshold, representing the required minimum confidence on predictions  for that class.
We detail our approach next, focusing first on a single class (Algorithm \ref{algo:cls}), and then generalizing it to multiple classes (Algorithm \ref{algo:cls2}).

\begin{algorithm*}[t!]
\small
\caption{Define when to stop computing probability of object $\mathbf{\hat{s}}_{d}$ belonging to  class $D_i$,  based on minimum confidence $\theta_i$, and minimum and maximum monitoring periods $\gamma_i$ and $\gamma^{max}$.}
\label{algo:cls}
\begin{algorithmic}[1]
     \Function{PerClassProb}{$\mathbf{\hat{s}}_{d}$, $\mathbf{c}_{D_i}$, $\theta_i$, $\gamma_i$, $\gamma^{max}$}
   \State{$p \gets0$}
    \State{$t_r \gets \gamma_i - 1$} \Comment{Start at previous window}
\While{$p < \theta_i$} \Comment{Extend monitoring period}
        \State{$t_r\gets t_r + 1$} \Comment{Move to next current window}
	\If {$ t_r > \gamma^{max}$} \Comment {Monitoring period ended}
		\State \Return {$\gamma^{max}, 0$} 
	\EndIf
         \State{$p \gets AlignComputeProb(\mathbf{\hat{s}}_{d}, \mathbf{c}_{D_i}, \theta_i,t_r)$}
    \EndWhile
    \State \Return{$t_r, p$} \Comment{Return monitoring period and probability}
    \EndFunction

\Function{AlignComputeProb}{$\mathbf{\hat{s}}_{d}$, $\mathbf{c}_{D_i}$, $\theta_i$,$t_r$}
           \State{$t_s \gets 1; \, p \gets 0 $}
            \While{ ($t_s \leq |\mathbf{c}_{D_i}| - t_r)  \quad  \text{and}  \quad (p < \theta_i$)}

\Comment {Iterate over  possible values of $t_s$, aligning both series}
            \State{$p' \propto exp(-dist(\mathbf{\hat{s}}_{d}^{[1:t_r]}, \mathbf{c}_{D_i}^{[t_s: t_s + t_r - 1]}))$}
            \State{$p \gets max(p, p')$}
            \State{$t_s \gets t_s + 1$}
        \EndWhile
	\State         \Return {$p$}
\EndFunction
\end{algorithmic}
\end{algorithm*}

Algorithm~\ref{algo:cls} shows how we define when to stop computing the probability  for a given class $D_i$.
The algorithm takes as input the object stream $\mathbf{\hat{s}}_{d}$, the class centroid $\mathbf{c}_{D_i}$,    the minimum confidence $\theta_i$ required to state that a new object belongs to $D_i$, as well as $\gamma_i$ and  $\gamma^{max}$,  the minimum   and maximum  thresholds for the monitoring period. The former is used to avoid computing distances with too few  windows, which may lead  to very high (but unrealistic) probabilities. 
The latter is used to guarantee that the algorithm ends. We allow different values of $\gamma_i$ and $\theta_i$ for each class as different popularity trends have overall different dynamics, requiring different thresholds\footnote{Indeed, initial experiments showed that using the same values of $\gamma_i$  (and $\theta_i$)   for all classes produces worse results.}.   The algorithm outputs the  number of monitored windows $t_r$ and the estimated probability $p$.
The loop in line 4 updates the stream with new observations (increases $t_r$), and function $AlignComputeProb$ computes the probability for a given $t_r$ by   trying all possible alignments (i.e., all possible values of $t_s$). For a fixed alignment (i.e., fixed $t_r$ and  $t_s$), $AlignComputeProb$ computes the distance   between both time series (line 15) and  the probability of $\mathbf{\hat{s}}_{d}$ belonging to $D_i$ (line 16).  It returns the largest probability representing the best alignment between $\mathbf{\hat{s}}_{d}$ and $\mathbf{c}_{D_i}$, for the given $t_r$ (lines 17 and 20).  Both loops that iterate over $t_r$ (line 4) and $t_s$ (line 15) stop when the probability exceeds the minimum  confidence $\theta_i$. The algorithm also stops when the monitoring period $t_r$ exceeds $\gamma^{max}$ (line 7), returning  a probability equal to $0$  to indicate that it was not possible to state the $\mathbf{\hat{s}}_{d}$  belongs to $D_i$ within the maximum monitoring period allowed ($\gamma^{max}$).

\begin{algorithm*}[t!]
\small
\caption{Define  when to stop computing probabilities for each object in $D^{test}$, considering the centroids of all classes ($\mathbf{C}_D$),  per-class minimum confidence  ($\bm{\theta}$) and monitoring period ($\bm{\gamma}$), and maximum monitoring period ($\gamma^{max}$).}
\label{algo:cls2}
\begin{algorithmic}[1]
    \Function{MultiClassProbs}{ $D^{test}$, $\mathbf{C}_D$, $\bm{\theta}$, $\bm{\gamma}$,$\gamma^{max}$}
    \State{$\mathbf{t} = [0]$} \Comment{Per-object monitoring period vector}
    \State{$\mathbf{P} = [[0]]$} \Comment{Per-object, per-class probability matrix}
    \State{$n_{objs} \gets |D^{test}|$ }\Comment{Number of objects to be monitored}
    \State{$t_r \gets min(\bm{\gamma})$} \Comment{Init $t_r$ with minimum $\gamma_i$}
    \While{$(t_r  \leq \gamma^{max}) \quad \text{and} \quad (n_{objs} > 0)$} 
         \ForAll{$\mathbf{\hat{s}}_d \in D^{test}$}  \Comment{Predict class for each object}
	      \ForAll {$\mathbf{c}_{D_i} \in \mathbf{C}_D$}  \Comment{Get  centroid of each class}
		   \State{$p^{[i]} \gets AlignComputeProb(\mathbf{\hat{s}}_{d},\mathbf{c}_{D_i}, \theta_i, t_r$)}
                \EndFor
                \State{$maxp \gets max(\mathbf{p})$}  \Comment{Get max. probability and corresponding class for  $t_r$}
                \State{$maxc \gets argmax(\mathbf{p})$} 
                \If{ $(maxp > \bm{\theta}^{[maxc]})$ \, \text{and} \, $(t_r \geq \bm{\gamma}^{[maxc]})$ } \Comment{Stop if maxp  and $t_r$ exceeds per-class thresholds}
                       \State{$\mathbf{t}^{[d]} \gets t_r$}  \Comment{Save current $t_r$}
                       \State{$\mathbf{P}^{[d]} \gets \mathbf{p}$}  \Comment{Save current $\mathbf{p}$ in row $d$}
		     \State{$n_{objs} \gets n_{objs} - 1$}
		     \State{$D^{test} \gets D^{test} - \{\mathbf{\hat{s}}_d\}$} 
                \EndIf
          \EndFor
	\State{$t_r \gets t_r + 1$}
      \EndWhile
\State \Return{$\mathbf{t}, \mathbf{P}$} \Comment{Return monitoring periods and probabilities}
    \EndFunction
\end{algorithmic}
\end{algorithm*}

We now extend Algorithm \ref{algo:cls} to compute probabilities and monitoring periods for all object streams in $D^{test}$, considering all  classes extracted from $D^{train}$.    Algorithm \ref{algo:cls2} takes as input the test set $D^{test}$, a matrix $\mathbf{C}_{D}$ with the class centroids,  vectors $\bm{\theta}$ and $\bm{\gamma}$ with per-class parameters, and $\gamma^{max}$. It outputs a vector $\mathbf{t}$ with the required monitoring period for each object, and a matrix $\mathbf{P}$ with the probability estimates for each object (row) and class (column), both initialized with 0 in all elements. Given a valid monitoring period $t_r$ (line 6), the algorithm monitors  each object  $d$ in $D^{test}$ (line 7) by first computing the probability of $d$ belonging to each class (line 9). It then takes, for each object $d$, the largest of the computed  probabilities  (line 11) and the associated class  (line 12), and tests whether it is possible to state that $d$ belongs to that class with enough confidence at  $t_r$, i.e., whether:  (1) the probability exceeds the minimum confidence for the class, {\it and} (2) $t_r$ exceeds the per-class minimum threshold (line 13).  If the test succeeds, the algorithm stops monitoring the object (line 16), saving the current $t_r$ and the per-class probabilities computed at this window in  $\mathbf{t}$ and $\mathbf{P}$ (lines 14-15).   After exhausting all possible monitoring periods ($t_r > \gamma^{max}$) or whenever the number of  objects being monitored $n_{objs}$ reaches 0, the algorithm returns. At this point, entries with $0$ in $\mathbf{P}$ indicate objects  for which no prediction was possible within the maximum monitoring period allowed  ($\gamma^{max}$). 

Having $\mathbf{P}$, a simple classifier can be built  by choosing  for each object (row) the class (column) with maximum probability. The value in $\mathbf{t}$ determines how early this classification can be done. 
However, we here employ a different strategy, using matrix $\mathbf{P}$ as input features to another classifier, as discussed below. We compare our proposed approach against the aforementioned simpler strategy in  Section \ref{sec:res}.

\subsubsection{Probabilities as Input Features to a Classifier}

Instead of directly extracting classes from $\mathbf{P}$, we choose to use this matrix as input features to another classification algorithm, motivated by previous results on the effectiveness of using  distances as  features to  learning methods~\citep{Coates2012}.  Specifically, we employ an extremely randomized trees classifier~\citep{Geurts2006}, as it has been  shown to be effective  on different datasets~\citep{Geurts2006}, requiring little or no pre-processing, besides producing models that can be more easily interpreted, compared to other  techniques like Support Vector Machines\footnote{We also used SVM learners, achieving  similar results.}.  Extremely randomized trees tackle the over fitting problem of more common decision tree algorithms by training a large ensemble of trees. They work as follows:  1)~for each node in a tree, the algorithm selects the best features for  splitting based on a random subset of all features; 2) split values are chosen at random. The decision of these trees are then averaged out to perform the final classification. Although feature search and split values are based on randomization, tree nodes are still chosen based on the maximization of some measure of discriminative power such as Information Gain, with the goal of improving classification effectiveness.

We  extend the set of probability features taken from $\mathbf{P}$ with  other features associated with the objects.  The set of  object features used depends on the type of UGC under study and characteristics of the datasets ($D$). We here use the  features shown in Table \ref{tab:feats}, which are further discussed in Section \ref{sec:data}, combining them with the probabilities in $\mathbf{P}$. We refer to this approach as {\it TrendLearner}.

Before continuing, we briefly discuss other strategies to combine classifiers as we have done. We experimented with these methods, finding them to be unsuitable to our dataset due to various reasons. For instance, we implemented Co-Training~\citep{Nigam2000}, a traditional semi-supervised label propagation approach. However, it failed to achieve better results than just combining the features, most likely because it depends on feature independence, which may not hold in our case. We also experimented with Stacking~\citep{Dzeroski2004},  which yielded similar results as the proposed  approach.  Nevertheless, either strategy might  be more effective on different datasets or types of UGC, an analysis that we leave for future work. 

\subsection{Putting It All Together} \label{sec:together}

A key point that remains to be discussed is how to define the input parameters of the  trend extraction approach,  that is, the number of trends $k$,  as well as the parameters of TrendLearner, namely  vectors $\bm{\theta}$ and $\bm{\gamma}$, $\gamma^{max}$, and the parameters of the  adopted classifier. 

We choose the number of trends $k$ based primarily on the $\beta_{CV}$ quality metric~\citep{Menasce2002}.  Let the intraclass distance be the distance between a time series and its centroid (the trend), and the interclass distance be the distance between different trends. The general purpose of the trend extraction is to minimize the variance of the intraclass distances while maximizing the variance of the interclass distances. The $\beta_{CV}$ is defined as the ratio of the coefficient of variation\footnote{The ratio of the standard deviation to the mean.} (CV) of intraclass distances to the CV of the interclass distances. The value of $\beta_{CV}$ should be computed for increasing values of $k$. The smallest $k$ after which the $\beta_{CV}$ remains roughly stable should be chosen~\citep{Menasce2002}, as a stable $\beta_{CV}$ indicates that new splits affect  only marginally the variations of  intra and interclass distances, implying that a well formed trend has been split.

Regarding the TrendLearner parameters, we here choose to constrain $\gamma^{max}$ with the maximum number of points  in our time series (100 in our case, as discussed in Section \ref{sec:data}). 
As for  vector parameters $\bm{\theta}$ and $\bm{\gamma}$,  a traditional cross-validation in the training set  to determine their optimal values  would  imply in a search over an exponential space of values. Moreover, note that it is fairly simple to achieve best classification results by setting $\bm{\theta}$ to all zeros and $\bm{\gamma}$ to large values, but this would lead to very late predictions (and possibly low remaining interest in the content after prediction). Instead, we suggest an alternative approach.  
Considering each class $i$ separately, we run a one-against-all  classification for objects of $i$ in $D^{train}$ for values of $\gamma_i$ varying from 1 till $\gamma^{max}$. We select the smallest value of $\gamma_i$ for which the  performance exceeds a minimum target (e.g., classification above random choice, meaning Micro-F1 greater than 0.5), and set $\theta_i$  to the average probability computed for all class $i$ objects for the selected $\gamma_i$. We repeat the same process for all classes. 
 Depending on the required tradeoff between prediction accuracy and remaining fraction of views, different performance targets could be used.
Finally, we use cross-validation in the training set to choose the parameter values for  the extremely randomized trees classifier, as further discussed in Section \ref{sec:res}.

\begin{algorithm*}
\small
\centering
\caption{Our Solution: Trend Extraction and Prediction}
\label{algo:tl}
\begin{algorithmic}[1]
    \Function{TrendExtraction}{$D^{train}$}
        \State{$k \gets 1$}
        \While{$\beta_{CV}$ is not stable}
            \State{$k \gets k + 1$}
            \State{$\mathbf{C}_D \gets KSC(D^{train}, k)$}
        \EndWhile
        \State{Store centroids in $\mathbf{C}_D$}
    \EndFunction
    \Function{TrendLearner}{$\mathbf{C}_D$, $D^{train}$, $D^{test}$}
    \State{$\bm{\theta}, \bm{\gamma}, \mathbf{P}^{train} \gets LearnParams(D^{train}, \mathbf{C}_D)$}
        \State{$TrainERTree(D^{train}, \mathbf{P}^{train} \bigcup$ obj. feats)} 
        \State{$\mathbf{t}, \mathbf{P} \gets MultiClassProbs(D^{test}, \mathbf{C}_D, \bm{\theta}, \bm{\gamma})$}
        \State \Return{$\mathbf{t},PredictERTree(D^{test}, \mathbf{P}\bigcup$ obj. feats)}
    \EndFunction
\end{algorithmic}
\end{algorithm*}

\begin{figure*}[t]
\centering
\centering
\begin{tikzpicture}
[node distance = 1cm, auto, font=\footnotesize,
every node/.style={node distance=1cm},
comment/.style={rectangle, inner sep=5pt, text width=4cm, node distance=0.25cm, font=\scriptsize},
force/.style={rectangle, draw, fill=black!10, inner sep=5pt, text width=1.5cm, text badly centered, minimum height=.25cm, font=\scriptsize}]

\node [force, fill=blue!20] (trends) {TrendExtraction (KSC)};
\node [below=.15cm of trends] (dummy) {};
\node [force, dashed, below=.5cm of trends, fill=green!20] (ugcpop) {Pop. Time Series (train)};
\node [force, right=.3cm of trends, fill=blue!20] (lp) {LearnParams \\ $\quad$};
\node [force, below=.5cm of lp, fill=blue!20] (lc) {TrainClassifier (ERTree)};
\node [force, dashed, below=.5cm of lc, fill=green!20] (dtrain) {Obj. Features (train)};
\node [force, right=.3cm of lp, fill=blue!20] (lr) {MultiClass \\ Probs};
\node [force, below=.5cm of lr, fill=blue!20] (pred) {UseClassifier \\ (ERTree)};
\node [force, dashed, right=.3cm of pred, fill=red!20] (objtest) {Obj. Features (test)};
\node [force, dashed, right=.3cm of lr, fill=red!20] (ugcstr) {Pop. Time Streams (test)};
\node [force, below=.5cm of pred, fill=white!20] (results) {Prediction \\ Results};
\node [rectangle, draw, thick, dotted, right=1cm of dummy, text width=4.1cm, text height=2.3cm, align=right, font=\scriptsize] (bb) {};
\node [above=-.1cm of bb.north, font=\scriptsize, align=right] (lb) {TrendLearner};

\path[->,thick] 
(ugcpop) edge (trends)
(trends) edge (lp)
(lp) edge (lc)
(lp) edge (lr)
(dtrain) edge (lc)
(ugcstr) edge (lr)
(lc) edge (pred)
(lr) edge (pred)
(objtest) edge (pred)
(pred) edge (results);

\draw[->,thick] (trends.north) -- ++(0,.7cm) -| node [near start] {} (lr.north);

\end{tikzpicture} 
\caption{Pictorial Representation of Our Solution}
\label{fig:sol}
\end{figure*}
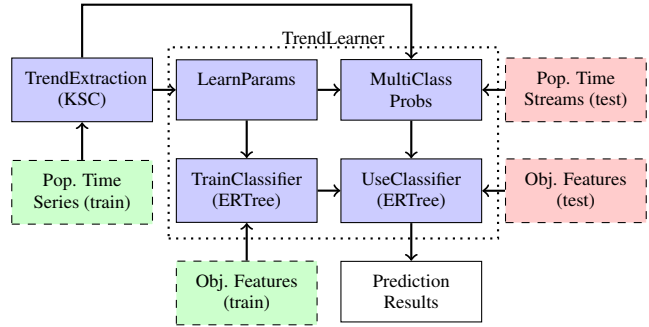

We summarize our solution to the early trend prediction problem in Algorithm~\ref{algo:tl}. In particular, TrendLearner works by first learning   the best parameter values and the classification model  from the training set ($LearnParams$ and $TrainERTrees$), and then applying the learned model to classify test objects ($PredictERTrees$), taking the class membership probabilities ($MultiClassProb$)  and other object features as inputs. A pictorial representation  is shown in Figure~\ref{fig:sol}. Compared to previous efforts \citep{Chen2013}, our method incorporates multiple classes, uses only centroids to compute class membership probabilities (which reduces time complexity),  and combines these probabilities with other object  features as inputs to a classifier, which, as  shown in Section \ref{sec:res}, leads to better results.

\section{Evaluation Methodology} \label{sec:methodology}


This section  presents  the metrics  (Section \ref{sec:metrics}) and datasets  (Section \ref{sec:data}) used in our evaluation.


\subsection{Metrics} \label{sec:metrics}

As discussed in Section \ref{sec:meth}, an inherent challenge of the early popularity trend prediction problem is to properly address the tradeoff between prediction accuracy and how early the prediction is made.  Thus, we  evaluate our method with respect to these two aspects.

We estimate prediction accuracy using the {\it standard} Micro and Macro $F1$ metrics,  which are computed from precision and recall. The precision of class $c$, $P(c)$, is the fraction of correctly classified videos out of those assigned to $c$ by the classifier, whereas the recall of class $c$, $R(c)$, is the fraction of correctly classified objects out of those that actually belong to that class. The $F1$ of class $c$ is given by: 
$ F1(c) = \frac{2 \cdot P(c) \cdot R(c)}{P(c) + R(c)}.$
 Macro F1 is the average $F1$ across all classes, whereas Micro F1 is computed from global precision and recall, calculated for all classes.


To complement the standard metrics above, we propose the use of novel metrics that we define to measure the effectiveness of the early predictions extracted by TrendLearner. These metrics are by no means replacements for standard classification evaluation metrics (such as the F1 defined above). That is, given that TrendLearner aims to capture the trade-off between accuracy and early predictions, our proposed novel metrics need to be evaluated together with the traditional ones. Recall that, our objectives are to evaluate both the: (1) accuracy of the classification; and (2) the possible loss of user interest in objects over time. 


We evaluate how early our correct  predictions are made computing  the remaining interest ($RI$) in the content {\it after} prediction. The $RI$ for an object $\mathbf{s}_d $  is defined as the fraction of all views up to a certain point in time (e.g., the day when the object was collected) that are received {\it after}  the prediction. That is, 
    $RI(\mathbf{s}_d, \mathbf{t}) = \frac{sum(\mathbf{s}_d^{[\mathbf{t}^{[d]}+1:n]})}{sum(\mathbf{s}_d^{[1:n]})}$
where $n$ is the number of points in $d$'s time series,   
 $t^{[d]}$ is the prediction time (i.e., monitoring period) produced by our method for $d$, 
and   function $sum$   adds up the elements of the input vector. In essence, this metric captures the future potential audience of   $\mathbf{s}_d $  after prediction.

We also  assess whether there is any bias in our {\it correct} predictions towards more (less) popular objects by computing  the 
 correlation  between  the total popularity and the remaining interest after prediction for each object.  A low correlation implies no bias,  while a strong positive (negative) correlation implies a bias towards earlier predictions for more (less) popular objects. We argue that, if any bias exists,  a bias towards more popular objects is preferred, as it implies larger remaining interests for those objects.   We use  both the Pearson linear correlation coefficient ($\rho_p$) and the Spearman's rank correlation coefficient ($\rho_s$) \citep{Jain1991}, as the latter does not assume linear relationships,  taking the logarithm of the total popularity  first due to the great skew in their distribution \citep{Figueiredo2011,Crane2008,Cha2009}.







\subsection{Datasets} \label{sec:data}

As case study, we focus on YouTube videos and use twodatasets, analyzed in  \citep{Figueiredo2011} and publicly available\footnote{\url{http://vod.dcc.ufmg.br/traces/youtime/}}. The \textbf{Top} dataset consists of 27,212 videos from the various top lists maintained by YouTube (e.g., most viewed and most commented videos), and the \textbf{Random topics} dataset includes  24,482 videos collected as results of  random queries submitted to YouTube's API\footnote{We do not claim this dataset is a random sample of YouTube videos. Nevertheless, for the sake of simplicity, we use the term Random videos to refer to videos from this dataset.}.

\begin{table}[ht]
\normalsize
\centering
\caption{Summary of Features}
\begin{tabular}{lll} \toprule
Class & Feature Name & Type  \\
\midrule
\multirow{2}{*}{Video}    & Video category           & Categorical   \\
                          & Upload date              & Numerical   \\
			& Video age & Numerical \\
		    & Time window size ($w$)   & Numerical    \\ \midrule
\multirow{2}{*}{Referrer}     & Referrer first date      & Numerical    \\
                          & Referrer \# of views & Numerical   \\ \midrule
\multirow{5}{*}{Popularity} & \# of views              & Numerical   \\
                          & \# of comments           & Numerical     \\
                          & \# of favorites          & Numerical     \\
                          & change rate of views   & Numerical     \\
                          & change rate of comments   & Numerical     \\
                          & change rate of favorites   & Numerical     \\
                          & Peak fraction         & Numerical     \\
\bottomrule
\label{tab:feats}
\end{tabular}
\end{table}

For each video, the datasets contain the following features (shown in Table~\ref{tab:feats}): the time series of the numbers of views, comments and favorites, as well as the ten most important referrers (incoming links), along with the date that referrer was first encountered,  the video's upload date and its category. The original datasets contain videos of various ages, ranging from days to years. We choose to study only videos with more than 100 days for two reasons. First, these videos tend to have their long term time series popularity more stable. Second, the KSC algorithm requires that all time series vectors $\mathbf{s}_d$   have the same dimension $n$. Moreover, the popularity time series provided by YouTube contains at most 100  points, independently of the video's age. Thus, by focusing only on videos with at least 100 days of age, we can use $n$ equal to 100 for all videos. After filtering younger videos out,  we were left with 4,527 and 19,562 videos in the Top and Random datasets, respectively.

\begin{table*}[t!]
\normalsize
\centering
    \caption{Summary of analyzed datasets}
    \begin{tabular}{l*{1}{c}c*{2}{c}} \toprule
        & \multicolumn{2}{c|}{Top} & \multicolumn{2}{c}{Random} \\
        \cmidrule(r){2-5}
        & $\mu$ & $\sigma$ & $\mu$ & $\sigma$ \\
        \cmidrule(r){2-5}
        \# of Views & 4,022,634 & 9,305,996 & 141,413 & 1,828,887 \\
        Video Age (days) & 632 & 402 & 583 & 339 \\
        Window $w$ (days) & 6.38 & 4.06 & 5.89 & 3.42 \\
        \hline
    \end{tabular}
    \label{tab:data}
\end{table*}

Table~\ref{tab:data} summarizes our two datasets, providing mean $\mu$ and standard deviation $\sigma$ for the number of views, age (in days), and time window duration $w$\footnote{$w$ is equal to the video age divided by 99 as the first point in the time series corresponds to the day before the upload day.}. Note that both average and median window durations are around or below one week. This is important as previous work~\citep{Borghol2011} pointed out that effective popularity growth models can be built based on weekly views.

\section{Experimental Results} \label{sec:res}

In this section, we present our results of our trend extraction (Section \ref{sec:trendres}) and  trend prediction (Section \ref{sec:predres}) approaches. 
We also show how TrendLearner can be used to improve the accuracy of state-of-the-art  popularity prediction models (Section \ref{sec:application}).
These results were computed using 5-fold cross validation, i.e., splitting the dataset $D$ into 5 folds, where 4 are used as training set $D^{train}$ and one as test set $D^{test}$, and rotating the folds  such that each fold is used for testing once.  As discussed in Section \ref{sec:ourapproach}, trends are extracted from $D^{train}$ and predicted for videos in $D^{test}$. 

Since we are dealing with time series, one might argue that a temporal split of the dataset into folds would be preferred to a random split, as we do here. However, we choose a random split because of the following.  Regarding the object features used as input to the prediction models, no temporal precedence is violated, as the features are computed only during the monitoring period $t_r$, {\it before} prediction.  All remaining features are based on the distances between the popularity curve of the object until $t_r$ and the class centroids (or trends).  As we argue below, the same trends/centroids found in our experiments were  consistently found in various subsets of each dataset,  covering various periods of time. Thus,  we expect the results to remain similar if a temporal split is done. However, a temporal split of  our dataset would require interpolations in the time series, as all of them have exactly 100 points regardless of video age. Such interpolations, which are not required in a random split, could introduce serious inaccuracies and compromise our analyses.


\subsection{Trend Extraction} \label{sec:trendres}


Recall that we used the $\beta_{CV}$ metric to determine the number of trends $k$  used by the KSC algorithm. In both datasets, we found $k$ to be stable after $4$ trends. We also checked centroids and class members for larger values of $k$, both visually and using other metrics (as in \citep{Yang2011}),  finding no reason for choosing  a different value\footnote{A possible reason would be the appearance of a new distinct class, which did not happen.}. Thus, we set $k=4$.  We also analyzed the centroids in all training sets, finding that the same 4 shapes appeared in every set.  Thus, we manually aligned classes based on their centroid shapes in different training sets so that class $i$ is the same in every set.  We also found  that, in  95\% of the cases, a video was always assigned to the same class in different sets. 

\begin{figure*}[t]
    \centering
    \includegraphics[scale=1.5]{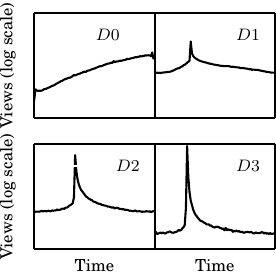}
    \caption{Popularity Trends in YouTube Datasets.}
    \label{fig:cents}
\end{figure*}

\begin{table*}[ht]
    \normalsize
    \centering
    \captionof{table}{Summary of popularity trends (classes)}
    \begin{tabular}{l*{4}{c}} \toprule
        & $D_0$ & $D_1$ & $D_2$ & $D_ 3$ \\
        \midrule
        & \multicolumn{4}{c}{Top Dataset} \\
        \cmidrule(r){2-5}
        \% of Videos & 22\% & 29\% & 24\% & 25\% \\
         Avg. \# of Views   & 711,868 & 6,133,348 & 1,440,469 & 1,279,506 \\
        Avg. Change Rate  in \# Views & 1112 & 395 & 51 & 67 \\
        Avg. Peak Fraction & 0.03 & 0.04 & 0.19 & 0.40  \\
        \midrule
        & \multicolumn{4}{c}{Random Dataset} \\
        \cmidrule(r){2-5}
        \% of Videos  & 21\% & 34\% & 26\% & 19\%\\
        Avg. \# of Views & 305,130 & 108,844 & 64,274 & 127,768\\
        Avg. Change Rate  in \# Views & 47 & 7 & 4 & 4\\
        Avg. Peak Fraction & 0.03 & 0.03 & 0.08 & 0.28\\
        \bottomrule
    \label{tab:summclus}
    \end{tabular}
\end{table*}

Figure~\ref{fig:cents} shows the popularity trends discovered in the Random dataset. Similar trends were also extracted from the Top dataset. Each graph shows the number of views as function of time, omitting scales as centroids are shape and volume invariants. The y-axes are in log scale to highlight  the importance of the peak.  We note that the KSC algorithm consistently produced the same popularity trends  for various randomly selected samples of the data, which are also consistent with similar shapes identified in other datasets \citep{Crane2008,Yang2011}.  We also note that the 4 identified trends might not perfectly match the popularity curves of {\it all} videos, as there might be  variations within each class.  However, our goal is not to perfectly model the popularity evolution of all videos. Instead, we aim at capturing the most prevalent trends, respecting time shift and volume invariants, and using them to improve popularity prediction. As we show in Section \ref{sec:application}, the identified trends can greatly improve state-of-the-art prediction models. 

Table~\ref{tab:summclus} presents, for each class, the percentage of videos belonging to it, as well as the average number of views, average change rate\footnote{Defined by the average of $p_{d,i+1} - p_{d,i}$ for each video $d$ represented by vector    $ \mathbf{s}_d = <p_{d,1}, p_{d,2}, \cdots, p_{d,n}>$.}, and average fraction of views at the  peak time window of these videos.
Note that class $D_0$ consists of videos that remain popular over time, as indicated by the large positive  change rates, shown in Table \ref{tab:summclus}. This behavior is specially strong in the Top dataset, with an average change rate of 1,112 views per window,  which corresponds to roughly a week  (Table \ref{tab:data}). Those videos also have no significant popularity peak, as the average fraction of views in the peak window is very small  (Table \ref{tab:summclus}). The other three classes are predominantly defined by a single popularity  peak, and are distinguished by the rate of decline after the peak: it is  slower in $D_1$, faster in $D_2$, and very sharp in $D_3$. These classes also exhibit very small change rates, indicating stability after the peak.

We also measured the distribution of different types of referrers and video categories across classes in each dataset. Under a Chi-square test with significance of $.01$, we found that 
the distribution differs from that computed for the aggregation of all classes, implying that these  features are somewhat  correlated with the class, and motivating their use to improve trend classification.

\subsection{Trend Prediction}
\label{sec:predres}

We now discuss our trend prediction results, which are averages of 5 test sets along with corresponding 95\% confidence intervals. We start by showing results that support our approach of computing class membership probabilities using only centroids as opposed to all class members, as  in \citep{Chen2013} (Section \ref{sec:resshapelets}). We then evaluate our  TrendLearner method, comparing it with three alternative approaches  (Section \ref{sec:resTrendLearner}).

\begin{table}[t!]
    \normalsize
    \centering
    \caption{ Classification using only centroids vs. using all class members: averages and 95\% confidence intervals.}
    \begin{tabular}{l*{1}{c}c|*{2}{c}} \toprule
   Monitoring     & \multicolumn{2}{c}{Centroid} & \multicolumn{2}{c}{Whole Training Set} \\
        \cmidrule(r){2-5}
  period $t_r$      & Micro F1 & Macro F1 & Micro F1 & Macro F1 \\
        \cmidrule(r){2-5}
        1 window    & $.24\pm.01$ & $.09\pm.00$ & $.29\pm.04$ & $.11\pm.01$ \\
        25 windows   & $.56\pm.02$ & $.52\pm.01$ & $.53\pm.04$ & $.44\pm.08$ \\
        50 windows   & $.67\pm.03$ & $.65\pm.03$ & $.64\pm.05$ & $.57\pm.09$ \\
        75 windows   & $.70\pm.02$ & $.68\pm.02$ & $.69\pm.08$ & $.61\pm.12$ \\
        \bottomrule
    \end{tabular}
    \label{tab:centsall}
\end{table}

\subsubsection{Are shapelets better than a reference dataset?} \label{sec:resshapelets}

We here discuss how the use of  centroids to compute class membership probabilities (Equation \ref{eq:prob}) compare to using all class members \citep{Chen2013}. For the latter,  the probability of an object belonging to a class is proportional to a summation over the exponential of the (negative) distance between the object  and every member of the given class.  

An important benefit of our approach is a reduction in running time: for a given object, it requires computing the distances to only $k$ time series, as opposed to the complete training set $|D^{train}|$, leading to a reduction in running time by a factor of $\frac{|D^{train}|}{k}$, as discussed in Section \ref{subsec:trendextraction}.
We here focus on the classification effectiveness of the probability matrix $\mathbf{P}$ produced by both approaches. To that end, we consider  a classifier that assigns the class with largest probability to each object, for both matrices.

Table \ref{tab:centsall} shows Micro and Macro F1 results for both approaches, computed for fixed monitoring periods $t_r$ (in number of windows) to facilitate comparison. We show results only for the Top dataset, as they are similar for the Random dataset.  Note that, unless the monitoring period is very short ($t_r$=1), both strategies produce  statistically tied results, with 95\% confidence. Thus, given the reduced time complexity,  using centroids only is more cost-effective. When using a single window {\it both} approaches are worse than random guessing (Macro F1 $=$ 0.25),  and thus are not interesting.

\subsubsection{TrendLearner Results} \label{sec:resTrendLearner}

We now compare  our {\bf TrendLearner}  method with three other trend prediction methods, namely:
(1) {\bf $\mathbf{P}$ only}: assigns the class with largest probability in $\mathbf{P}$ to an object; (2) {\bf $\mathbf{P}$ + ERTree}: trains an extremely randomized trees learner using $\mathbf{P}$ only as features; (3) {\bf ERTree}: trains an extremely randomized trees learner using only the object features in Table~\ref{tab:feats}. Note that  TrendLearner combines  ERTree and $\mathbf{P}$ + ERTree.  Thus, a comparison of these four methods allows us to assess the benefits of combining both sets of features.

For {\it  all}   methods,  when classifying a video $d$, we only consider features of that video available up until $\mathbf{t}^{[d]}$,  the time window  when TrendLearner stopped monitoring $d$.
We also use  the same best values for parameters shared by the methods, chosen as discussed in Section \ref{sec:together}. Both Tables \ref{tab:params-top} (for the Top dataset) and~\ref{tab:params-rnd} (for the Random dataset), show the best values of  vector parameters $\bm{\gamma}$ and $\bm{\theta}$,  selected considering a Macro-F1 of at least 0.5 as performance target (see   Section \ref{sec:together}). These results are averages across all training sets, along with  95\% confidence intervals. The variability is low in most cases, particular for $\bm{\theta}$.
 Recall that $\gamma^{max}$ is set to 100. Regarding the extremely randomized trees classifier, we set the size of the ensemble to 20 trees, and the feature selection strength equal to the square root of the total number of features, common choices for this classifier~\citep{Geurts2006}. We then apply cross-validation within the training set to choose the smoothing length parameter ($n_{min}$), considering values equal to  $\{1, 2, 4,$ $8, 16, 32\}$. We refer  to \citep{Geurts2006} for more details on the parametrization of extremely randomized trees. 

\begin{table}[t!]
    \normalsize
    \centering
    \caption{Best values for vector parameters $\bm{\gamma}$ and $\bm{\theta}$ (averages and 95\% confidence intervals) for the Top dataset}
    \begin{tabular}{lcccc} \toprule
        & \multicolumn{4}{c}{Top Dataset}\\
        \cmidrule(r){2-5}
        & $D_0$ & $D_1$ & $D_2$ & $D_3$ \\
        \cmidrule(r){2-5}
        $\bm{\theta}$& $.250\pm.015$ & $.257\pm.001$ & $.272\pm.003$ & $.303\pm.006$ \\
        $\bm{\gamma}$& $28\pm16$ & $89\pm8$ & $5\pm0.9$ & $3\pm0.5$ \\
        \bottomrule
    \end{tabular}
    \label{tab:params-top}
\end{table}
        
\begin{table}[t!]
    \normalsize
    \centering
    \caption{Best values for vector parameters $\bm{\gamma}$ and $\bm{\theta}$ (averages and 95\% confidence intervals) for the Random dataset}
    \begin{tabular}{lcccc} \toprule
        & \multicolumn{4}{c}{Random Dataset} \\
        \cmidrule(r){2-5}
        & $D_0$ & $D_1$ & $D_2$ & $D_3$ \\
        \cmidrule(r){2-5}
        $\bm{\theta}$& $.250\pm.001$ & $.251\pm.001$ & $.269\pm0.001$ & $.317\pm0.001$\\
        $\bm{\gamma}$& $33\pm0.6$ & $74\pm2$ & $45\pm9$ & $17\pm3$\\
        \bottomrule
    \end{tabular}
    \label{tab:params-rnd}
\end{table}

Still analyzing Tables~\ref{tab:params-top} and~\ref{tab:params-rnd}, we note that classes with smaller peaks ($D_0$ and $D_1$) need longer minimum monitoring periods $\gamma_i$, likely because even small fluctuations may be confused as peaks
due to the scale invariance of the  distance metric used (Equation \ref{eq:dist})\footnote{Indeed, most of these videos are wrongly classified into  either $D_2$ or $D_3$ for shorter monitoring periods.}. However, after this period, it is somewhat easier  to determine whether the object belongs to one of those classes (smaller values of $\theta_i$). In contrast, classes with higher peaks ($D_2$ and $D_3$) usually require shorter monitoring periods, particularly in the Top dataset, where videos have popularity peaks with larger fractions of views (Table \ref{tab:summclus}).  Indeed, by cross-checking results in Tables \ref{tab:summclus}, \ref{tab:params-top} and \ref{tab:params-rnd}, we find that classes with smaller fractions of videos in the peak window ($D_0$ and $D_1$ in Top, and $D_0$, $D_1$ and $D_2$ in Random) tend to require longer minimum monitoring periods so as to avoid confusing small  fluctuations with peaks from  the other classes.

\begin{table}[t]
    \normalsize
    \centering
    \caption{Comparison of trend prediction methods for both datasets (averages and 95\% confidence intervals) for the Top dataset}
        \begin{tabular}{lcccc} 
        \toprule
        & \multicolumn{4}{c}{Top Dataset}\\
        \cmidrule(r){2-5}
        & $\mathbf{P}$ only & $\mathbf{P}$+ERTree & ERTree & TrendLearner \\
        \cmidrule(r){2-5}
        Micro F1    & $.48\pm.06$ & $.48\pm.06$ & $.58\pm.01$ & $.62\pm.01$ \\
        Macro F1    & $.44\pm.06$ & $.44\pm.06$ & $.57\pm.01$ & $.61\pm.01$ \\
        \bottomrule
    \end{tabular}
    \label{tab:restl-top}
\end{table}
\begin{table}[t]
    \normalsize
    \centering
    \caption{Comparison of trend prediction methods for both datasets (averages and 95\% confidence intervals) for the Random dataset}
    \begin{tabular}{lcccc} 
        \toprule
        & \multicolumn{4}{c}{Random Dataset} \\
        \cmidrule(r){2-5}
        & $\mathbf{P}$ only & $\mathbf{P}$+ERTree & ERTree & TrendLearner \\
        \cmidrule(r){2-5}
        Micro F1    & $.67\pm.02$ & $.62\pm.01$ & $.65\pm.01$ & $.71\pm.01$\\
        Macro F1    & $.69\pm.02$ & $.63\pm.01$ & $.63\pm.01$ & $.70\pm.01$\\
        \bottomrule
    \end{tabular}
    \label{tab:restl-rnd}
\end{table}

We now discuss our classification results, focusing first on the Micro and Macro F1 results, shown in Table \ref{tab:restl-top} and Table~\ref{tab:restl-rnd}, for the Top and Random datasets respectivelly. From both tables we can see that TrendLearner consistently outperforms all other methods in both datasets and on both metrics, except for Macro F1 in the Random dataset, where it is statistically tied with the second best approach ($\mathbf{P}$  only).  In contrast, there is no clear winner among the other three methods across both datasets. Thus,  combining  probabilities and object features   brings clear benefits over using either set of features separately. For example, in the Top dataset, the gains over the alternatives in average Macro F1 vary from 7\% to 38\%, whereas the average improvements in Micro F1  vary from 7\% to 29\%.  Similarly, in the Random dataset, gains in average Micro and Macro F1 reach up to 14\% and 11\%, respectively.  Note that TrendLearner performs somewhat better in the Random dataset, mostly because videos in that dataset are monitored for longer, on average (larger values of $\gamma_i$). However, this superior results comes with a reduction in remaining interest after prediction, as we  discuss below. 

We note that the joint use of both probabilities and object features renders TrendLearner more robustness
to some (hard-to-predict) videos. Recall that, as discussed in Section \ref{subsec:prob}, Algorithm \ref{algo:cls2} may, in some cases, return a probability equal to  $0$ to indicate that a prediction was not possible within the maximum monitoring period allowed. Indeed, this happened for  1\% and 10\% of the videos in the Top and Random datasets, respectively, which have  popularity curves that do not closely follow  any of the extracted trends.  The  results for
the   $\mathbf{P}$ only   and   $\mathbf{P}$ + ERTree  methods  shown in Tables \ref{tab:restl-top} and \ref{tab:restl-rnd}   do {\it not} include such videos, as these methods are  not able to do predictions for them (since they rely only on the probabilities). However,  both   ERTree  and TrendLearner are able to perform predictions for such videos by exploiting the object features, since at least the video category and upload date are readily available as soon as the video is posted. Thus,  the results of these  two methods in Tables \ref{tab:restl-top} and \ref{tab:restl-rnd} contemplate the predictions for {\it all} videos\footnote{For the cases with probability equal to $0$, the predictions of TrendLearner and   ERTree  were made with $t_r$=$\gamma^{max}$, when Algorithm \ref{algo:cls2} stops. Since we set $\gamma^{max}$=100, those predictions were made at the last time window, using all available information to compute object features. Nevertheless, note that, in those cases, the remaining interest ($RI$) after prediction is equal to 0.}.

\begin{figure*}[t!]
 \centering
 \mbox{\subfigure[Remaining Interest (RI)  ]{\includegraphics[width=0.3\linewidth]{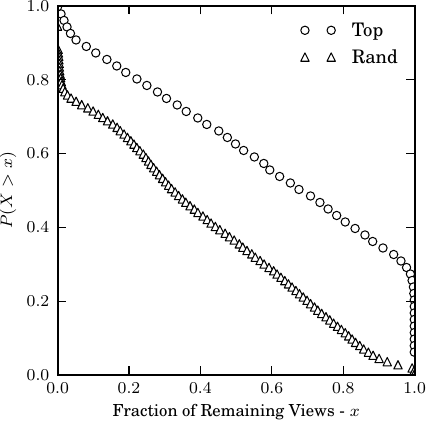}}}\hfill
 \mbox{\subfigure[Total Views vs. RI (Top)]{\includegraphics[width=0.3\linewidth]{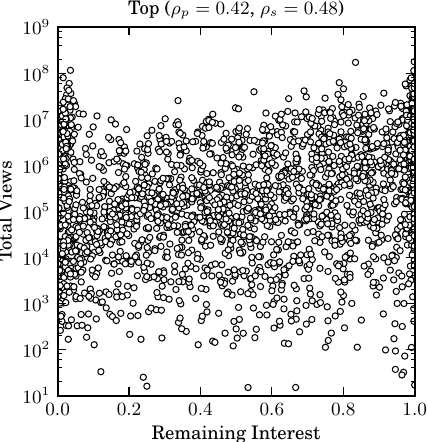}}}\hfill
 \mbox{\subfigure[Total View vs. RI (Random)]{\includegraphics[width=0.3\linewidth]{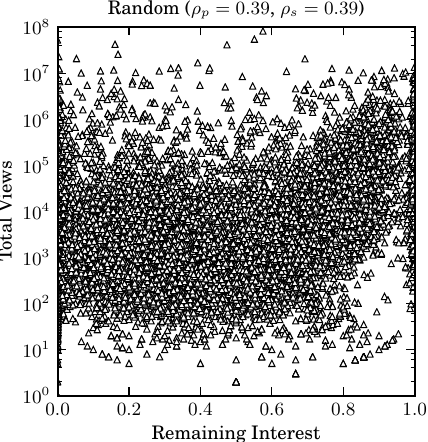}}}
 \caption{Remaining Interest (RI) and Correlations Between Popularity and RI for Correctly Classified Videos.}
 \label{fig:early}
\end{figure*}

We now turn to the other side of the tradeoff and discuss how early the predictions are made.  These results are the same for all four aforementioned methods as all of them use the  prediction time returned by TrendLearner.  For all {\it correctly} classified videos, we report the remaining interest $RI$ after prediction, as well as the Pearson ($\rho_p$) and Spearman ($\rho_s$) correlation coefficients   between remaining interest and  (logarithm of) total popularity (i.e., total number of views),  as informed in our datasets.  

Figure \ref{fig:early}(a) shows the complementary cumulative distribution of the fraction of $RI$ after prediction for both datasets, while Figures \ref{fig:early}(b) and \ref{fig:early}(c)  (log scale on the y-axis) show the  total number of views and the $RI$ for each video  in the Top and Random datasets, respectively.  All three graphs were produced for the union of the videos in all test sets.  Note that, for 50\% of the videos, our predictions are made before at least 68\% and 32\% of the views are received, for  Top and Random videos, respectively. The same $RI$ of at least 68\% of views is achieved for 21\% of videos in the Random dataset. In general, for a significant number of videos in both datasets, our correct predictions are made before a large fraction of their views are received, particularly in the Top dataset.  

We also point out a  great variability in the duration of the monitoring periods produced by our solution: while only a few windows are required for some videos, others have to be monitored for a longer period. Indeed, the coefficients of variation of these monitoring periods  are 0.54 and 1.57 for the Random and Top datasets, respectively. This result emphasizes the need for choosing a monitoring period on a per-object basis, a novel aspect of our approach, and not  use the same fixed value.

Moreover, the scatter plots in Figures \ref{fig:early}(b-c)  show that some moderately positive correlations  exist between the total number of views and $RI$. Indeed, 
$\rho_p$ and $\rho_s$ are equal to 0.42 and 0.48, respectively, in the Top dataset, while both metrics are equal to 0.39  in the Random dataset.
Such results imply that our solution is somewhat biased towards more popular objects, although the bias is not very strong.  In other words,  for more popular videos, TrendLearner is able to produce accurate predictions by  potentially  observing a smaller fraction of their total views, in comparison with less popular videos. This is a nice property, given that such predictions can drive advertisement placement and content replication/organization decisions which are concerned mainly with the most popular objects.

\subsection{Applicability to Regression Models} \label{sec:application}

Motivated by results in~\citep{Yang2010,Pinto2013}, which showed that knowing popularity trends {\em beforehand} can improve the accuracy of regression-based popularity prediction models, we here assess whether our trend predictions are good enough for that purpose. To that end, we use the state-of-the-art ML and MRBF regression models proposed in \citep{Pinto2013}. The former is a multivariate linear regression model that uses   the popularity  acquired  by the object $d$ on each time window up to a reference date $t_r$ (i.e., $p_{d,i}$, $i=1...t_r$) to  predict its popularity at a  target date $t_t = t_r+\delta$.  The latter extends the former by including features based on Radial Basis Functions (RBFs) to  measure the similarity between $d$ and   specific examples, previously selected from the training set.  

Our goal is to evaluate whether our trend prediction results can improve these  models. 
Thus, as  in \citep{Pinto2013}, we use the mean Relative Squared Error (mRSE) to assess the prediction accuracy of
  the ML and MRBF models in two settings: (1) a general model, trained using the whole dataset (as in \citep{Pinto2013}); (2)  a specialized model, trained for each {\it predicted} class. For the latter, we first use  our solution to predict the   trend of a video. We then train ML and MRBF models considering as reference date each value of $\mathbf{t}^{[d]}$ produced by TrendLearner for each video $d$. Considering a prediction lag $\delta$ equal to 1, 7, and 15, we measure the mRSE of the predictions for target date $t_t = \mathbf{t}^{[d]} + \delta$. 
\begin{table*}[t!]
    \normalsize
    \centering
    \caption{Mean Relative Squared Error  Various Prediction Models and  Lags $\delta$  (averages and 95\% confidence intervals)}
        \begin{tabular}{lcccccc} \toprule
       Prediction Model & \multicolumn{3}{c}{Top Dataset} & \multicolumn{3}{c}{Random Dataset} \\
        \cmidrule{2-7}
        & $\delta=1$ & $\delta=7$ & $\delta=15$ &$\delta=1$ & $\delta=7$ & $\delta=15$ \\ \cmidrule{2-7}

        generalML               & $.09\pm.005$ & $.42\pm.02$ & $.75 \pm.04$ & $.01\pm.001$ &  $.06\pm.005 $ & $.11\pm.01$\\ 
        generalMRBF             & $.08\pm.005$ & $.52\pm.05$ & $1.29\pm.17$ & $.01\pm.001$ &  $.1\pm.01   $ & $.26\pm.03$\\  
        best SSM                & $.76\pm.01 $ & $.63\pm.02$ & $.64 \pm.03$ & $.90\pm.002$ &  $.69\pm.005 $ & $.54\pm.006$\\   
        specializedML           & $.08\pm.005$ & $.27\pm.01$ & $.38 \pm.02$ & $.009\pm.001$ & $.04\pm.0003$ & $.06\pm.003$\\  
        specializedMRBF         & $.08\pm.005$ & $.32\pm.04$ & $.47 \pm.08$ & $.009\pm.001$ & $.04\pm.0004$ & $.06\pm.008$\\  
\bottomrule
    \end{tabular}
    \label{tab:specialized}
\end{table*}
We also compare our specialized models against the  state-space models (SSMs) proposed in \citep{Radinsky2012}.  These models are  variations of a basic state-space Holt-Winters model that represent query and click frequency in Web search, capturing various aspects of popularity dynamics (e.g., periodicity,  bursty behavior, increasing trend). All of them take  as input the popularity time series during the monitoring period $t_r$. Thus, though originally proposed for the  Web search domain, they can be directly applied to our context. 
Both regression and state-space models are parametrized as originally proposed\footnote{The only exception is the number of  examples used to compute similarities in the MRBF model: we used 50 examples, as opposed to the suggested 100 \citep{Pinto2013}, as it led to better results in our datasets.}. 

Table \ref{tab:specialized} shows average mRSE for each model along with  95\% confidence intervals, for all datasets and prediction lags. 
Comparing our specialized models and the original ones they build upon, we find that using our solution to build trend-specific models greatly improves prediction accuracy, particularly for larger values of $\delta$. The reductions in mRSE vary from 10\% to 77\% (39\%, on average) in the Random dataset, and  from 11\% to 64\% (33\%, on average) in the Top dataset\footnote{The only exception is  the MRBF model for $\delta$=$1$ in the Top dataset, where  general and specialized models produce tied results.}.  The specialized models also greatly outperform the state-space models: the reductions in mRSE over the best state-space model  are at least 89\% and 27\% in the Random and Top datasets (94\% and 59\%, on average).  These results offer strong indications of the usefulness of our trend predictions for predicting popularity measures.

Finally, it is important to discuss why the state-space models did not work well in our context. The main reason we found was that Holt-Winters based models can only capture the linear trends in time series, that is, linear growth and decay. By using the KSC distance function, we can identify and group UGC time series with non-linear trends~\citep{Matsubara2012,Yang2011}, and create specific prediction models for these cases. Also, these models are trained independently for each target object, using early points of the time series. Another possible reason for the low performance in our context might be that, unlike in \citep{Radinsky2012} where the models were trained with hundreds of points of each time series,  we here use much less data (only points   up  to $\mathbf{t}^{[d]}$).

\section{Conclusions} \label{sec:disc}

In this article, we have identified and formalized a new research problem. To the extent of our knowledge, we are the first work to tackle the problem of {\it early prediction} of popularity trends in UGC. We were motivated in studying this problem based on our previous knowledge on the complex patterns and causes of popularity in UGC~\citep{Figueiredo2011}. Different from other kinds of content, e.g., news, which have clear definitions of monitoring periods, target and prediction dates for popularity, the complex nature of UGC calls for a popularity prediction solution which is able to determine these dates automatically. We here provided such a solution -- TrendLearner. 

We have also proposed  a novel two-step learning approach for early prediction of popularity trends of UGC. Moreover, we defined new metrics for measuring the effectiveness of popularity of UGC content, the remaining interest, which is optimized by TrendLearner as to provide not only accurate, but also timely, predictions. Thus, unlike previous work, we addresses the tradeoff between prediction accuracy and remaining interest in the content after prediction on a per-object basis. 
        
We performed an extensive experimental evaluation of our method, comparing it with state-of-the-art, representative solutions of the literature. Our experimental results on two YouTube datasets  showed that our method not only outperforms other approaches for trend prediction (a gain of up to 38\%) but also achieves such results before 50\% or 21\% of videos (depending on the dataset) accumulate more than 32\% of their views, with a slight bias towards earlier predictions for more popular videos.  Moreover, when applied jointly with recently proposed regression based models to predict the popularity of a video at a future date, our method outperforms state-of-the-art regression and state-space based models, with gains in accuracy of at least 33\% and 59\%, on average, respectively.

As future work, we plan to further investigate  how different types of UGC (e.g., blogs and Flickr photos) differ in their popularity evolution as well as which factors (e.g., referrers, content quality) impact this evolution. 

\section*{Acknowledgments} 
This research is partially funded by the Brazilian National  Institute of Science and Technology for  Web Research  (MCT/CNPq/INCT Web Grant Number 573871/2008-6), and by the authors' individual grants from Google, CNPq, CAPES and Fapemig.




\bibliographystyle{abbrv}
\bibliography{bibs}

\begin{thebibliography}{10}

\bibitem{Ahmed2013}
M.~Ahmed, S.~Spagna, F.~Huici, and S.~Niccolini.
\newblock {A Peek Into the Future: Predicting the Evolution of Popularity in
  User Generated Content}.
\newblock In {\em Proc. WSDM}, 2013.

\bibitem{Borghol2011}
Y.~Borghol, S.~Mitra, S.~Ardon, N.~Carlsson, D.~Eager, and A.~Mahanti.
\newblock {Characterizing and Modeling Popularity of User-Generated Videos}.
\newblock {\em Performance Evaluation}, 68(11):1037--1055, 2011.

\bibitem{Castillo2014}
C.~Castillo, M.~El-Haddad, J.~Pfeffer, and M.~Stempeck.
\newblock {Characterizing the life cycle of online news stories using social
  media reactions}.
\newblock In {\em Proc. CSCW}, 2014.

\bibitem{Cha2009}
M.~Cha, H.~Kwak, P.~Rodriguez, Y.-Y. Ahn, and S.~Moon.
\newblock {Analyzing the Video Popularity Characteristics of Large-Scale User
  Generated Content Systems}.
\newblock {\em IEEE/ACM Transactions on Networking}, 17(5):1357--1370, 2009.

\bibitem{Chen2013}
G.~H. Chen, S.~Nikolov, and D.~Shah.
\newblock {A Latent Source Model for Nonparametric Time Series Classification}.
\newblock In {\em Proc. NIPS}, 2013.

\bibitem{Coates2012}
A.~Coates and A.~Ng.
\newblock {Learning Feature Representations with K-Means}.
\newblock {\em Neural Networks: Tricks of the Trade}, pages 561--580, 2012.

\bibitem{Crane2008}
R.~Crane and D.~Sornette.
\newblock {Robust Dynamic Classes Revealed by Measuring the Response Function
  of a Social System}.
\newblock {\em Proceedings of the National Academy of Sciences},
  105(41):15649--53, 2008.

\bibitem{Duong2013}
Q.~Duong, S.~Goel, J.~Hofman, and S.~Vassilvitskii.
\newblock {Sharding social networks}.
\newblock In {\em Proc. WSDM}, Feb. 2013.

\bibitem{Dzeroski2004}
S.~D\v{z}eroski and B.~\v{Z}enko.
\newblock {Is Combining Classifiers with Stacking Better than Selecting the
  Best One?}
\newblock {\em Machine Learning}, 54(3):255--273, 2004.

\bibitem{Figueiredo2014}
F.~Figueiredo, J.~Almeida, and M.~Gon\c{c}alves.
\newblock {Improving the Effectiveness of Content Popularity Prediction Methods
  using Time Series Trends}.
\newblock In {\em Proc. ECML/PKDD Predictive Analytics Challenge Workshop},
  2014.

\bibitem{Figueiredo2011}
F.~Figueiredo, F.~Benevenuto, M.~Gon\c{c}alves, and J.~Almeida.
\newblock {On the Dynamics of Social Media Popularity: A YouTube Case Study}.
\newblock {\em ACM Trans. Internet Technol.}, 14(4):24:1--24:23, 2014.

\bibitem{Geurts2006}
P.~Geurts, D.~Ernst, and L.~Wehenkel.
\newblock {Extremely Randomized Trees}.
\newblock {\em Machine Learning}, 63(1):3--42, 2006.

\bibitem{Gill2013}
P.~Gill, V.~Erramilli, A.~Chaintreau, B.~Krishnamurthy, D.~Papagiannaki, and
  P.~Rodriguez.
\newblock {Follow the Money: Understanding Economics of Online Aggregation and
  Advertising}.
\newblock In {\em Proc. IMC}, 2013.

\bibitem{Golbandi2013}
N.~G. Golbandi, L.~K. Katzir, Y.~K. Koren, and R.~L. Lempel.
\newblock {Expediting Search Trend Detection via Prediction of Query Counts}.
\newblock In {\em Proc. WSDM}, 2013.

\bibitem{Hu2014}
Q.~Hu, G.~Wang, and P.~S. Yu.
\newblock {Deriving Latent Social Impulses to Determine Longevous Videos}.
\newblock In {\em Proc. WWW}, 2014.

\bibitem{Jain1991}
R.~Jain.
\newblock {\em {The Art of Computer Systems Performance Analysis: Techniques
  for Experimental Design, Measurement, Simulation, and Modeling}}.
\newblock Wiley, 1991.

\bibitem{Jiang2014}
L.~Jiang, Y.~Miao, Y.~Yang, Z.~Lan, and A.~G. Hauptmann.
\newblock Viral video style: A closer look at viral videos on youtube.
\newblock In {\em Proc. ICMR}, ICMR '14, 2014.

\bibitem{Lee2010}
J.~G. Lee, S.~Moon, and K.~Salamatian.
\newblock {An Approach to Model and Predict the Popularity of Online Contents
  with Explanatory Factors}.
\newblock In {\em Proc. WIC}, volume~1, 2010.

\bibitem{Lerman2010}
K.~Lerman and T.~Hogg.
\newblock {Using a Model of Social Dynamics to Predict Popularity of News}.
\newblock In {\em Proc. WWW}, 2010.

\bibitem{Leskovec2011}
J.~Leskovec.
\newblock {Social Media Analytics}.
\newblock In {\em Proc. WWW}, 2011.

\bibitem{Li2013}
H.~Li, X.~Ma, F.~Wang, J.~Liu, and K.~Xu.
\newblock {On Popularity Prediction of Videos Shared in Online Social
  Networks}.
\newblock In {\em Proc. CIKM}, 2013.

\bibitem{Matsubara2012}
Y.~Matsubara, Y.~Sakurai, B.~A. Prakash, L.~Li, and C.~Faloutsos.
\newblock {Rise and Fall Patterns of Information Diffusion}.
\newblock In {\em Proc. KDD.}, 2012.

\bibitem{Menasce2002}
D.~Menasc\'{e} and V.~Almeida.
\newblock {\em {Capacity Planning for Web Services: Metrics, Models, and
  Methods}}.
\newblock Prentice Hall, 2002.

\bibitem{Nigam2000}
K.~Nigam and R.~Ghani.
\newblock {Analyzing the Effectiveness and Applicability of Co-training}.
\newblock In {\em Proc. CIKM}, 2000.

\bibitem{Pinto2013}
H.~Pinto, J.~Almeida, and M.~Gon\c{c}alves.
\newblock {Using Early View Patterns to Predict the Popularity of YouTube
  Videos}.
\newblock In {\em Proc. WSDM}, 2013.

\bibitem{Radinsky2012}
K.~Radinsky, K.~Svore, S.~Dumais, J.~Teevan, A.~Bocharov, and E.~Horvitz.
\newblock {Behavioral Dynamics on the Web: Learning, Modeling, and Prediction}.
\newblock {\em ACM Transactions on Information Systems}, 32(3):1--37, 2013.

\bibitem{Szabo2010}
G.~Szabo and B.~A. Huberman.
\newblock {Predicting the Popularity of Online Content}.
\newblock {\em Communications of the ACM}, 53(8):80--88, 2010.

\bibitem{Vakali2012}
A.~Vakali, M.~Giatsoglou, and S.~Antaris.
\newblock {Social Networking Trends and Dynamics Detection via a Cloud-Based
  Framework Design}.
\newblock In {\em Proc. WWW}, 2012.

\bibitem{Yang2010}
J.~Yang and J.~Leskovec.
\newblock {Modeling Information Diffusion in Implicit Networks}.
\newblock In {\em Proc. ICDM}, 2010.

\bibitem{Yang2011}
J.~Yang and J.~Leskovec.
\newblock {Patterns of Temporal Variation in Online Media}.
\newblock In {\em Proc. WSDM}, 2011.

\bibitem{Ye2011}
L.~Ye and E.~Keogh.
\newblock {Time Series Shapelets: A Novel Technique that Allows Accurate,
  Interpretable and Fast Classification}.
\newblock {\em Data Mining and Knowledge Discovery}, 22(1-2):149--182, 2011.

\bibitem{Yin2012}
P.~Yin, P.~Luo, M.~Wang, and W.-C. Lee.
\newblock {A Straw Shows Which Way the Wind Blows: Ranking Potentially Popular
  Items from Early Votes}.
\newblock In {\em Proc. WSDM}, 2012.

\bibitem{Yu2015}
H.~Yu, L.~Xie, and S.~Sanner.
\newblock {Exploring the Popularity Phases of YouTube Videos: Observations,
  Insights, and Prediction}.
\newblock In {\em Proc. ICWSM}, 2015.

\bibitem{Zeng2010}
D.~Zeng, H.~Chen, R.~Lusch, and S.-H. Li.
\newblock {Social Media Analytics and Intelligence}.
\newblock {\em IEEE Intelligent Systems}, 25(6):13--16, 2010.

\end{thebibliography}







\end{document}